\documentclass[twocolumn,floatfix,pra,showpacs]{revtex4}
\usepackage{graphicx}
\usepackage{amssymb}
\usepackage{amsmath}
\usepackage{color}
\usepackage{psfrag}
\usepackage{epsfig}
\usepackage{bbm}
\usepackage{bm}
\newcommand{\ket}[1]{| #1 \rangle}

\newcommand{\rb}[1]{\left( #1 \right)}

\newcommand{\ew}[1]{\langle #1 \rangle}
\newcommand{\beq}{\begin{eqnarray}}
\newcommand{\eeq}{\end{eqnarray}}

\newcommand{\op}[2]{| #1 \rangle \langle #2 |}

\newcommand{\eq}[1]{Eq.~(\ref{#1})}
\newcommand{\fig}[1]{Fig.~\ref{#1}}

\newcommand{\kett}[1]{| #1 \rangle\!\rangle }
\newcommand{\braa}[1]{\langle\!\langle #1|}
\newcommand{\eww}[1]{\langle\! \langle #1\rangle\! \rangle}
\newcommand{\opp}[2]{| #1 \rangle\! \rangle\langle\! \langle #2 |}
\begin{document}
\title{Three-level mixing and dark states in transport through quantum dots}
\author{Clive Emary, Christina P\"{o}ltl and Tobias Brandes}
\affiliation{
  Institut f\"ur Theoretische Physik,
  Hardenbergstr. 36,
  TU Berlin,
  D-10623 Berlin,
  Germany
}

\date{\today}
\begin{abstract}
  We consider theoretically the transport through the double quantum dot structure of the recent experiment of C.~Payette {\it et al.} [Phys.~Rev.~Lett.~{\bf 102}, 026808 (2009)] and calculate stationary current and shotnoise.
  Three-level mixing gives rise to a pronounced current suppression effect, the character of which charges markedly with bias direction.
  We discuss these results in connexion with the dark states of coherent population trapping in quantum dots.
\end{abstract}
\maketitle

\begin{figure}[tb]
  \begin{center}
    \epsfig{file=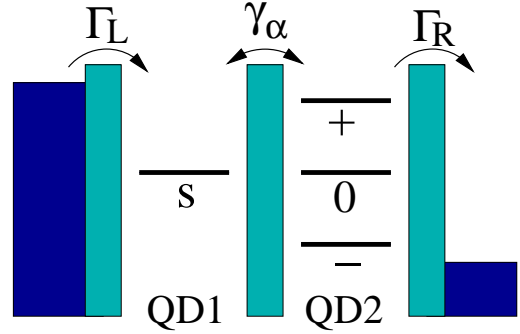, clip=true,width=0.8\linewidth}
    \caption{
      Double quantum dot with a bias window that includes the single probe $s$-level in QD1 and three levels of QD2. The depicted bias configuration is as in Ref.~\cite{pay09}, which we describe here as forward bias.  In the sequential tunneling regime, electron tunneling is described by the rates $\Gamma_L$ from left lead to QD1, $\Gamma_R$ from QD2 to the right lead, and by $\gamma_\alpha$; $\alpha=0,\pm$ between the dots.
    \label{FIG_sketch}
  }
  \end{center}
\end{figure}

In a recent experiment \cite{pay09}, Payette and co-workers studied the transport through a double quantum dot (DQD) in which the source-side QD (QD1) had a single electronic level within the transport window, whilst the drain-side dot (QD2) possessed three (see \fig{FIG_sketch}).  Gate voltages enabled the position of the former ``$s$-level'' to be adjusted and thus used as a probe of the second QD.
Due to non-ellipticity, the levels of QD2 were found not to be the familiar Fock-Darwin (FD) levels \cite{kou01}, but rather mixtures of them.  This gave rise to a distinctive feature 
in the tunneling magnetospectrum consisting of an avoided crossing with a central line running through it.
Strikingly, this central current line was not continuous as a function of magnetic field, as one might expect, but rather showed a strong suppression near the centre of the avoided crossing.  The authors of Ref.~\cite{pay09} suggested a connection between this phenomenon and that of the all-electronic coherent population trapping (CPT) of Refs.~\cite{mic06,gro06,CE07,pol09}.   It is the aim of this paper to explore this connexion further. 

We use a master equation treatment and calculate stationary current and shotnoise.  We consider a source-drain bias direction both as in Ref.~\cite{pay09} (forward bias), as well in the opposite direction (reverse bias).  Both bias directions yield a current suppression, but as our calculations here reveal, the character is rather different in each case.  In forward bias, the current suppression valley is wide (proportional to the mixing energy between the levels) as observed in the experiment of Ref.~\cite{pay09} and the shotnoise is subPoissonian.  In the reverse bias configuration, the current suppression valley is narrow (proportional to the coupling rate with the leads) and the current statistics are strongly superPoissonian.
We argue that only in the latter case does the current blocking mechanism bear strong resemblance to coherent population trapping.

\section{model}

We assume strong Coulomb blockade such that at most one excess electron can occupy the DQD at any one time and write the Hamiltonian of the complete system as
\beq
  H = H_1 + H_2 + H_{12} + H_\mathrm{leads} + V
 .
\eeq
The Hamiltonian of the first dot reads $H_1 = \epsilon_s \op{s}{s}$ with $\ket{s}$ denoting the single QD1 $s$-type orbital.  Denoting the bare FD levels in the second dot as $\ket{i};~i=1,2,3$, we take the Hamiltonian of the second dot to be of the form
\beq
  H_2 &=& 
  E_B(\op{1}{1}-\op{3}{3})
  \nonumber\\
  &&+T(\op{1}{2}+\op{2}{1}+\op{2}{3} +\op{3}{2} )
  ,
  \label{H2}
\eeq
where energy $E_B = c_B (B-B_0)$ with $c_B$ a constant, describes the magnetic field dependence of the FD levels (here assumed linear for levels 1 and 3 and constant for level 2), and $T$ is the coupling strength between the levels.  We assume that the coupling between the two dots can be described by
\beq
  H_{12} = \Omega\sum_{i=1}^{3}\rb{\op{s}{i}+\op{i}{s}}
  ,
\eeq
with common coupling parameter $\Omega$. 
Finally, Hamiltonian 
$H_\mathrm{leads} =\sum_{k.X} \varepsilon_{k X}c_{k X}^\dag c_{k X}$ 
describes two standard fermionic reservoirs ($X=L,R$: left, right), and
\beq
  V = 
  \sum_{k} 
  \rb{
  V_{kL} c_{k L}^\dag \op{0}{s}
  + \sum_{i=1}^3
  V_{k R} c_{k R}^\dag \op{0}{i} 
  }+ \mathrm{h.c.}
  \label{V}
  ,
\eeq 
with $\ket{0}$ the empty DQD state,  describes the coupling of the dots to the leads.
 Note that here we have chosen the simplest configuration of parameters --- our aim is the qualitative understanding of such systems and not the quantitative reproduction of the results of Ref.~\cite{pay09}.

The eigenstates of \eq{H2} play the determining role in the transport through the system;  we shall denote them $ \ket{\Psi_0}$ and $\ket{\Psi_\pm}$, corresponding to eigenenergies $\epsilon_0=0$ and $\epsilon_\pm = \pm \sqrt{E_B^2 + 2 T^2}$ such that $H_2 \ket{\Psi_\alpha}=\epsilon_\alpha \ket{\Psi_\alpha}$; $\alpha = 0,\pm$.  The most important of these three states is that belonging to eigenvalue zero:
\beq
  \ket{\Psi_0} &=& 
  \ket{\Psi_0(E_B)} = 
  \frac{\rb{
	 -T \ket{1}
	 + E_B \ket{2}
	  +T\ket{3}
	}}{\sqrt{E_B^2 + 2 T^2}}
	.
	\label{Psi0} 
\eeq

\section{Master equation}

All three barriers of the double quantum dot are high, and thus a treatment in terms of Fermi's golden rule is appropriate for all tunneling in the system. Furthermore, we assume large bias such that, for a given bias direction, tunneling to/from the leads is unidirectional, and all relevant Fermi functions are either zero or one.  We describe tunneling to/from left and right leads with the rates $\Gamma_L$ and $\Gamma_R$ respectively and define $\Gamma = \Gamma_L + \Gamma_R$.  From the form of \eq{V}, the right-lead rates are the same for all three FD orbitals, and thus also for all three eigenstates $\ket{\Psi_\alpha}$.

Tunnelling between the dots is governed by the matrix elements of $H_{12}$ and therefore by the overlaps between the states in the two dots.  Denoting the overlaps of the $s$-level with the QD2 FD states as $\ew{s|i}=s_i$; $i=1,2,3$, we have, for example,
\beq
  \ew{s|\Psi_0} = 
  \frac{1}{\sqrt{E_B^2 + 2 T^2}}
  \rb{
     T(s_3-s_1)
	 + E_B s_2
  }
  .
\eeq 
This eigenstate-overlap clearly vanishes for $E_B =   T(s_1-s_3)/s_2$.   In principle, overlaps $s_i$ must be determined from calculation with orbital wave functions.  However, here we make the simple assumption that all $s_i$ are the same. 
This is justified because the essential feature that $\ew{s|\Psi_0}$ vanishes remains regardless of the particular values of $s_i$.
We then set these overlaps to unity, since they can be subsumed into the rate $\gamma$, to be defined  below.
The squares of the relevant matrix elements are then 
\beq
   |\ew{s|\Psi_0}|^2 &=&	\frac{E_B^2}{E_B^2+2T^2}
   \label{MEs0}
   \\
   |\ew{s|\Psi_\pm}|^2 &=& 1 + \frac{T^2}{E_B^2+2T^2}
     \pm\frac{2T}{\sqrt{E_B^2+2T^2}}
   .
   \label{MEspm}
\eeq

Following \cite{spr04}, we then take the hopping rates between states $s$ and $\alpha$ to be
\beq
  \gamma_{\alpha}= \gamma  |\ew{s|\Psi_\alpha}|^2 
  L(|\epsilon_s - \epsilon_\alpha|, \Gamma)
  ;\quad \alpha = 0,\pm
  ,
\eeq
where we have assumed a Lorentzian broadening of the levels, $L(x,w)=(1+(2x/w)^2)^{-1}$, and $\gamma = \gamma(\Omega)$ sets the overall scale for these rates \cite{FN1}.

\begin{widetext}

With forward bias (as depicted in \fig{FIG_sketch}), the Liouvillian (rate matrix) of the system in a basis of populations of states 
(`empty', $s$, $\Psi_0$, $\Psi_-$ ,$\Psi_+$) reads
\beq
  {\cal L}_\mathrm{fwd} (\chi) =
  \rb{
  \begin{array}{ccccc}
    -\Gamma_L & 0 & \Gamma_R & \Gamma_R & \Gamma_R\\
    \Gamma_L e^{i \chi} & -\gamma_0 - \gamma_- - \gamma_+ 
        & \gamma_0 & \gamma_- & \gamma_+ \\
    0 & \gamma_0 &-\gamma_0 -\Gamma_R & 0 & 0\\
    0 & \gamma_- & 0  & -\gamma_- -\Gamma_R & 0 \\
    0 & \gamma_+ & 0  & 0  & - \gamma_+ -\Gamma_R \\
  \end{array}
  }.
\eeq
In reverse bias, the right chemical potential lies above all three levels in QD2, with that on the left lying below the QD1 $s$-level. The Liouvillian for this situation is 
\beq
  {\cal L}_\mathrm{rev} (\chi) =
  \rb{
  \begin{array}{ccccc}
    -3\Gamma_R & \Gamma_L e^{i \chi} & 0 & 0 & 0\\
    0 & -\Gamma_L -\gamma_0 - \gamma_- - \gamma_+ 
        & \gamma_0 & \gamma_- & \gamma_+ \\
    \Gamma_R & \gamma_0 &-\gamma_0  & 0 & 0\\
    \Gamma_R & \gamma_- & 0  & -\gamma_- & 0 \\
    \Gamma_R & \gamma_+ & 0  & 0  & - \gamma_+  \\
  \end{array}
  }
  .
\eeq
\end{widetext}
Here we have added counting field $\chi$ to facilitate the calculation of the current and shotnoise \cite{lev93,bag03}. The density matrix itself evolves under the action of the $\chi=0$ Liouvillian.  For example, in the forward bias case we have
$
  \dot{\rho }(t) = {\cal L}_\mathrm{fwd}(0) \rho(t)
$.

\subsection{Current statistics formalism}

The current statistics of our model can straightforwardly be calculated using the jump-super-operator formalism of full counting statistics \cite{fli04, CE09}.
The stationary density matrix of the system, written as vector $\kett{\rho_\mathrm{stat}}$ is defined by ${\cal L}(0) \kett{\rho_\mathrm{stat}}=0$.  The corresponding left ``Trace vector'' is $\braa{\phi_0} = (1,1,1,1,1)$ such that $\eww{\phi_0|\rho_\mathrm{stat}}=1$.
Let us define the stationary state ``expectation value'' $\eww{\ldots} = \eww{\phi_0|\ldots|\rho_\mathrm{stat}}$, jump super-operator ${\cal J} = \frac{d}{d(i\chi)} \left.{\cal L}(\chi) \right|_{\chi\to 0}$, and the pseudo-inverse propagator
${\cal R}(z) = {\cal Q}\left[z-{\cal L}(0)\right]^{-1}{\cal Q}$ 
with ${\cal Q}={ \mathbbm 1}-{\cal P}$ and ${\cal P} = \opp{\rho_\mathrm{stat}}{\phi_0}$.  In this notation, the current and zero-frequency shotnoise read
\beq
  \ew{I} &=& \eww{{\cal J}}
  \\
  S &=& \eww{{\cal J}} + 2 \eww{{\cal J}{\cal R}(0){\cal J}}
  .
\eeq 
We further define the Fano factor as $F=S/\ew{I}$.

\section{Results}

In all the following, we set $\Gamma_L = \Gamma_R$ for ease of presentation. Furthermore, in the experiment tunneling rates were much smaller that the level-mixing strength, $T \gg \Gamma_R, \gamma$, we will use this fact in various approximate results.

\subsection{Forward bias}
\begin{figure}[tb]
  \psfrag{ES}{\!\!\!\!$\epsilon_s/\Gamma_R$}
  \psfrag{EZ}{\!\!\!\!\!$E_B/\Gamma_R$}
  \psfrag{II}{\!\!\!\!$\ew{I}/\Gamma_R$}
  \psfrag{F}{$F$} 
  \psfrag{(a)}{(a)}
  \psfrag{(b)}{(b)}
  \begin{center}
   \epsfig{file=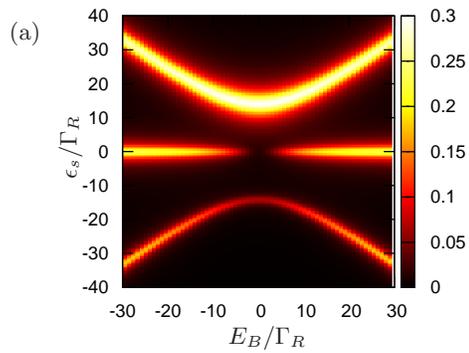, clip=true,width=\linewidth}
   \epsfig{file=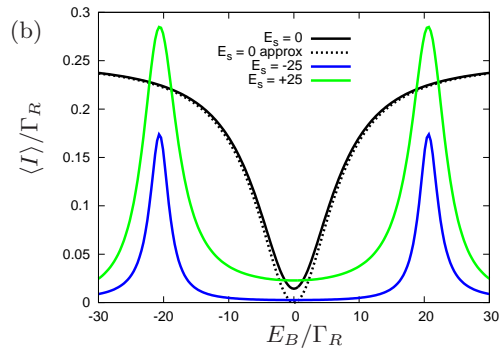, clip=true,width=0.77\linewidth}
    \caption{
       Current through the DQD in the forward bias configuration: 
       {\bf (a)} density plot as a function of magnetic energy $E_B$ and $s$-level position $\epsilon_s$.
       {\bf (b)} sections at $\epsilon_s = 0$ (black), $\epsilon_s = -25 \Gamma_R$ (blue), and $\epsilon_s = 25 \Gamma_R$ (green). Also plotted (dashed) is the approximate current of \eq{Ifwdapp} (for $\epsilon_s = 0$).
       A strong current suppression is observed around the point $\epsilon_s = E_B=0$, but note that a small current does flow at this point, however.
       Further parameters are: $\gamma=\Gamma_L=\Gamma_R$ and $T=10\Gamma_R$.
    \label{FIGIfwd}
  }
  \end{center}
\end{figure}

Figure \ref{FIGIfwd} shows the current through the system as a function of magnetic energy $E_B$ and QD1 level position, $\epsilon_s$.  The general structure of the measurements of Ref~\cite{pay09} --- an avoided crossing with a line through the middle --- is reproduced, with current suppression near $\epsilon_s = E_B = 0$ clearly present. Near this point, the $s$-level is close to resonance with the QD2 state $\ket{\Psi_0}$ and, if we ignore contributions from the other two levels, the current through the system may be approximated as
\beq
  \ew{I}_\mathrm{fwd} \approx 
  \frac{\gamma_0\Gamma_R }
  {\Gamma_R  + 3 \gamma_0}
  \label{Ifwdapp}
  .
\eeq
The rate $\gamma_0$ is proportional to the matrix element $|\ew{s|\Psi_0}|^2$ which, from \eq{MEs0} is seen to vanish at $E_B=0$.  Within this approximation, the stationary state of the system at $\epsilon_s=E_B=0$ is $\rho_\mathrm{stat} = \op{s}{s}$, with an electron trapped in the $s$-level due to the vanishing of the matrix element. In this approximation, the current at this point is zero. 
From \fig{FIGIfwd}b, however, it is clear that the current is not completely suppressed at $\epsilon_s=E_B=0$, but is finite due to the conduction through the other two states $\ket{\Psi_\pm}$.  This residual current can be estimated as
$I_\mathrm{fwd} \approx 3\gamma \Gamma_R^2/(2T^2)$, which need not be negligible.

\begin{figure}[tb]
  \psfrag{ES}{\!\!\!\!$\epsilon_s/\Gamma_R$}
  \psfrag{EZ}{\!\!\!\!\!$E_B/\Gamma_R$}
  \psfrag{II}{\!\!\!\!$\ew{I}/\Gamma_R$}
  \psfrag{F}{$F$}
  \psfrag{(a)}{(a)}
  \psfrag{(b)}{(b)}
  \begin{center}
    \epsfig{file=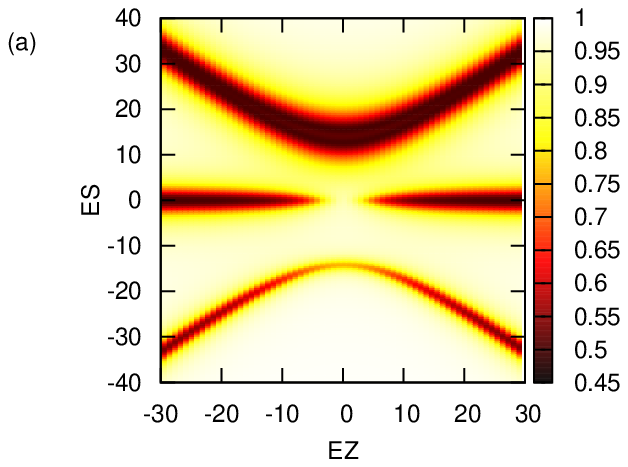, clip=true,width=\linewidth}
    \epsfig{file=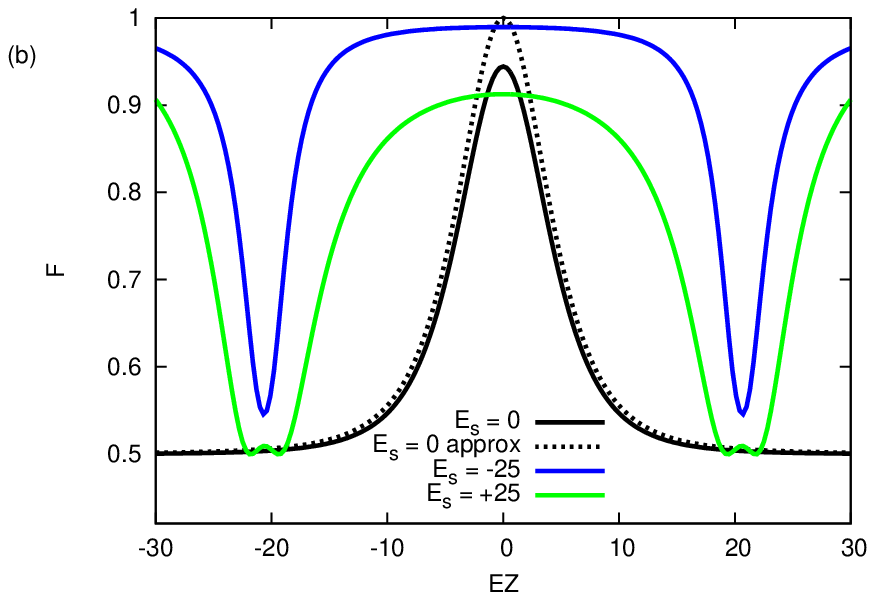, clip=true,width=0.77\linewidth}
    \caption{
      As \fig{FIGIfwd}, but here the shotnoise Fano factor is shown.  Along the resonant lines of high current, the shotnoise is subPoissonian with a Fano factor $F \approx 1/2$. Away from resonance, as well as around the central current suppression region, the shotnoise approaches the Poissonian value $F=1$ from below.
    \label{FIGFfwd}
  }
  \end{center}
\end{figure}
The width of current suppression feature at $\epsilon_s=0$ can be approximated as follows.  Close to $E_B=0$, $\gamma_0$ is small, and \eq{Ifwdapp} can be further approximated as $I_\mathrm{fwd} \approx \gamma_0$.  On the other hand, far from $E_B=0$, the current saturates to the constant value 
$I_\mathrm{fwd} \approx \Gamma_R\gamma/(3 \gamma + \Gamma_R)$. The value of the magnetic field at the point where these two behaviours cross can be found by setting the two limiting values equal, and solving for $E_B$. Equating this value to half the width of the current suppression valley we find
\beq
  w_\mathrm{fwd} = \sqrt{\frac{8 \Gamma_R}{3\gamma}} T
  ,
\eeq
which shows the width of the current suppression valley to be proportional to the level-mixing energy $T$.

The shotnoise Fano factor for this bias direction is shown in \fig{FIGFfwd}.  Especially evident is that the Fano factor is everywhere less then (or equal to) unity, corresponding to the familiar subPoissonian statistics of anti-bunched electron transfer.  Again assuming that only the central resonance determines the transport in the neighbourhood of the current suppression, we can approximate
\beq
  F_\mathrm{fwd} \approx 
  \frac{\Gamma_R^2 + 2 \Gamma_R \gamma_0 + 5 \gamma_0^2}
  {(\Gamma_R + 3 \gamma_0)^2}
  ,
\eeq
which is clearly always less than or equal to unity.  Along the central resonance ($\epsilon_s=0$), the Fano factor reaches a maximum value $F\approx 1-6\gamma \Gamma_R/T^2$ at $E_B=0$ and a limiting value of
$F \approx (5 \gamma^2+2\gamma \Gamma_R+\Gamma_R^2)/(\Gamma_R+3\gamma)^2$ for large $E_B$ along the $\epsilon_s=0$ line.

\subsection{Reverse bias}

\begin{figure}[tb]
  \psfrag{ES}{\!\!\!\!$\epsilon_s/\Gamma_R$}
  \psfrag{EZ}{\!\!\!\!\!$E_B/\Gamma_R$}
  \psfrag{II}{\!\!\!\!$\ew{I}/\Gamma_R$}
  \psfrag{F}{$F$}
  \psfrag{(a)}{(a)}
  \psfrag{(b)}{(b)}
  \begin{center}
    \epsfig{file=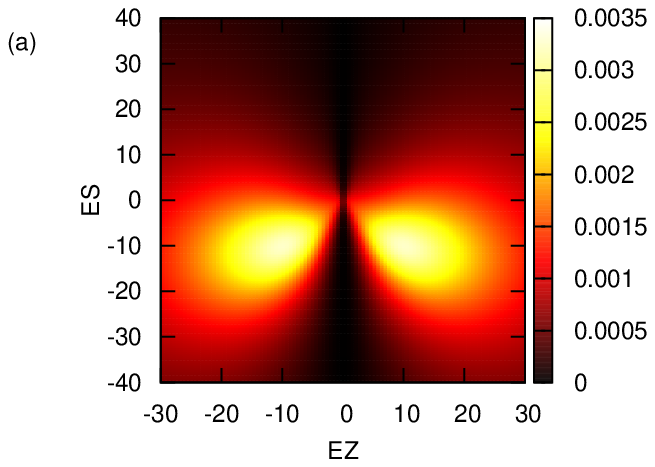, clip=true,width=\linewidth}
    \epsfig{file=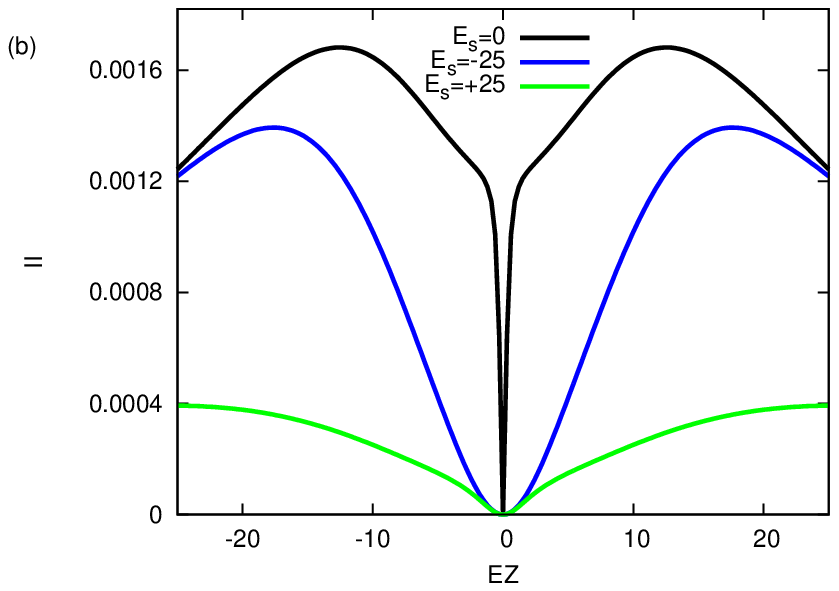, clip=true,width=0.77\linewidth}
    \caption{
      Current in the reverse bias, parameters as \fig{FIGIfwd}.  Here, the level structure of the second dot is not resolved. Rather, two large current peaks are observed. Along the $E_B=0$ axis, the current is completely suppressed, irrespective of the $s$-level position.  This is attributed to the formation of the CPT dark state of \eq{darkstate}.
    \label{FIGIrev}
  }
  \end{center}
\end{figure}

Figure \ref{FIGIrev} shows the current with the source-drain bias in the opposite direction.  Once again, the current shows a suppression at $E_B=0$, but unlike the forward bias case, this suppression extends for all positions of the $s$-level.  The second significant feature of this suppression is that the current is {\it exactly} zero at $E_B=0$, even for $T/\gamma$ finite.  It is easily shown that for $E_B=0$, the stationary density matrix of the system is
$
  \rho_\mathrm{stat} = \op{\Psi_0(0)}{\Psi_0(0)}
$,
which clearly shows that in the long time limit, an electron is trapped in the DQD in the pure state 
\beq
  \ket{\Psi_0(0)} = \frac{1}{\sqrt{2}}\rb{\ket{3}-\ket{1}}
  \label{darkstate}
\eeq

We can obtain approximate expressions for the current as follows.
If $\Gamma_R$ and $\gamma$ are of the same order-of-magnitude, then, for any choice of $\epsilon_s$, at least one of $\gamma_\alpha$ will be much smaller than $\Gamma_R$.  We can then write  the current as
\beq
  I_\mathrm{rev} \approx \frac{3\gamma_0 \gamma_+ \gamma_-}
  {\gamma_0\gamma_+ + \gamma_0\gamma_- + \gamma_+ \gamma_-}
  .
\eeq
Concentrating about the point $\epsilon_s=0$, we can say that near $E_B=0$ there is always a regime in which $\gamma_0$ is the smallest rate (matrix element disappears). In this case the current becomes
$
   I_\mathrm{rev} \approx 3\gamma_0 
$.  This expression describes the form on the sharp dip in the current about $E_B=0$.  Further away from $E_B=0$, $\gamma_0$ becomes the largest of the $\gamma_i$ rates since its matrix element returns to a typical non-suppressed value, and the other two rates are off-resonant. In this case, the current becomes 
$ I_\mathrm{rev} \approx 3 \gamma_+ \gamma_-/(\gamma_+ + \gamma_- )$.  An estimate of the width of the current-suppression feature can then be obtained from the cross over between these two behaviours which occurs when $\gamma_0 =\gamma_+$ (NB: for $\epsilon_s=0$, $\gamma_-<\gamma_+$). Solving for $E_B$, we find the width to be 
\beq
  w_\mathrm{rev} = \sqrt{6+4\sqrt{2}} \Gamma_R
  ,
\eeq
which is proportional to the broadening induced by the contacts.


\begin{figure}[tb]
  \psfrag{ES}{\!\!\!\!$\epsilon_s/\Gamma_R$}
  \psfrag{EZ}{\!\!\!\!\!$E_B/\Gamma_R$}
  \psfrag{II}{\!\!\!\!$\ew{I}/\Gamma_R$}
  \psfrag{F}{$F$}
  \psfrag{(a)}{(a)}
  \psfrag{(b)}{(b)}
  \begin{center}
    \epsfig{file=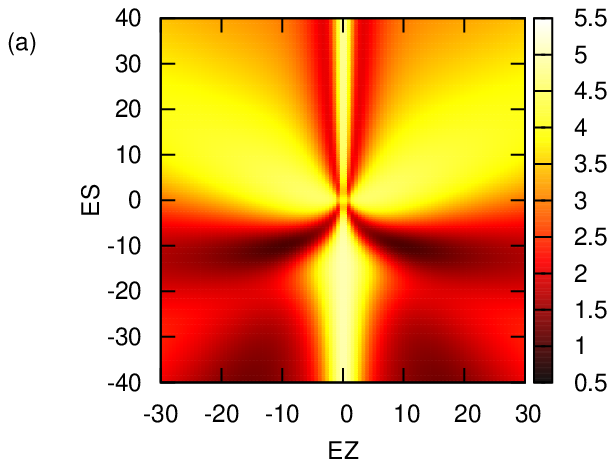, clip=true,width=\linewidth}
    \epsfig{file=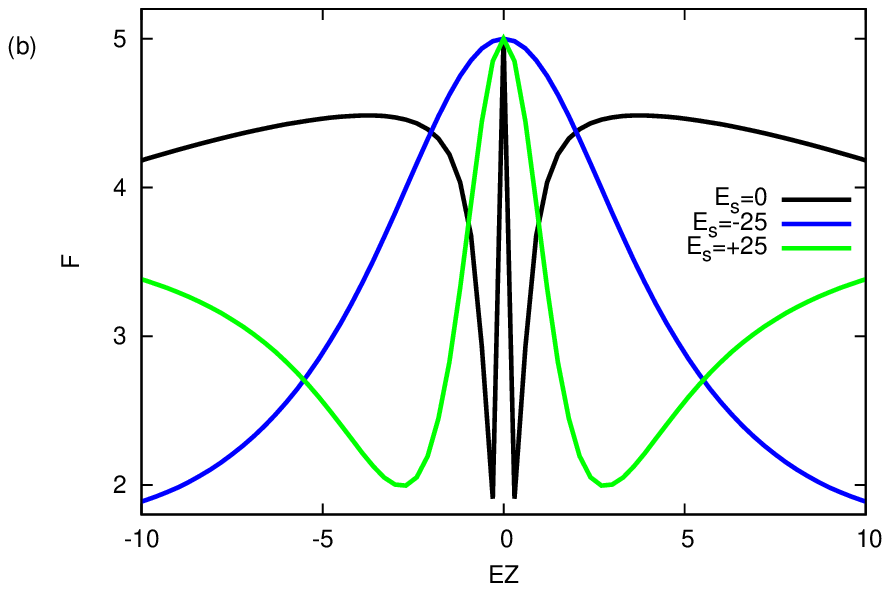, clip=true,width=0.77\linewidth}
    \caption{ 
      Shotnoise Fano factor in the reverse bias, parameters as \fig{FIGFfwd}.  In contrast to the forward bias case, the shotnoise is almost everywhere superPoissonian. For $E_B=0$, the Fano factor is exactly 5.
    \label{FIGFrev}
  }
  \end{center}
\end{figure}

The corresponding Fano factor is shown in \fig{FIGFrev}. In strong contrast to the forward bias case, the noise here is almost everywhere superPoissonian, and in particular in the neighbourhood of the current suppression.  Analytic expressions for $F$ are, in this case unwieldy.  However, without any further approximation, we find that at $E_B=0$ the Fano factor is simply $F=5$ independent of all further parameters.

\section{Discussion}

The foregoing results allow us to form a physical picture of the transport mechanisms at work in the current suppression here.

For forward bias, near the point $\epsilon_S=E_B = 0$, conduction comes through three channels which are weakly transmitting: two ($\pm$) on account of their distance in energy from resonance with the probe level at $\epsilon_s$, and one ($0$) on account of the vanishing of the matrix element for hopping between the two dots. 
In this case, the steady state of the DQD is approximately that of an electron trapped in the QD1 probe level $\rho_\mathrm{stat}\approx \op{s}{s}$.  This trapping is not exact, however, and a current still flows at $\epsilon_S=E_B = 0$ due to conduction through the off-resonant channels.
As a natural consequence of having a set of weakly transmitting channels, the statistics are subPoissonian.
This situation resembles somewhat the isospin blockade of Ref.~\cite{jac04}.

In contrast, the current in the reverse direction for $E_B=0$ is exactly zero --- not just for $\epsilon_s =0 $, but irrespective of probe-level position.
In this case, the dot electron is trapped in the state: $2^{-1/2}\rb{\ket{3}-\ket{1}}$.  This is a pure superposition state and is directly analogous to the dark state of the triple quantum dot CPT \cite{mic06}.  As in the triple dot case, the corresponding current statistics are superPoissonian.  This may be understood in terms of the dynamical channel blockade \cite{dcb}, since we have one weakly transmitting channel (that associated with the dark state), and two normally conducting ones.

A further distinguishing feature between these two blockade situations is that the width of the forward bias suppression valley is proportional to the mixing amplitude $T$ (the large energy scale in the model), whereas that in reverse bias is proportional to the lead-coupling rate $\Gamma_R$ (the small energy scale).
We also mention that the reverse-bias suppression is robust if we increase the interdot coupling $\Omega$, whereas the forward-bias feature washes out as the three resonances start to overlap.

The reverse-bias CPT effect described here should be more robust with regards to dephasing induced by e.g. background charge fluctuations than the dark state of the triple QD since it is substantially more localised.  No interdot coherence is required.  Furthermore, the ease in tuning the magnetic field to precisely locate the dark state, as compared with the gate voltages in the triple QD, makes this an excellent set-up for the further study of CPT and dark states in mesoscopic transport.

\begin{acknowledgements}
This work was supported by the WE Heraeus foundation and by DFG grant BR 1528/5-1.
\end{acknowledgements}



\begin{thebibliography}{99}
\bibitem{FN1}  We have repeated these calculations with a coherent treatment of the internal coupling, and for $T \gg \Gamma_{L,R}, \gamma$, the results are in good agreement with this simpler rate equation approach.
\end{thebibliography}
\end{document}